\documentclass[manuscript]{aastex}

\begin{document}
\pagenumbering{arabic}

\title{A COMPARISON BETWEEN THE HALF-LIGHT RADII, LUMINOSITIES, AND UBV COLORS OF GLOBULAR CLUSTERS IN M31 AND THE GALAXY}

\author{Sidney van den Bergh}
\affil{Dominion Astrophysical Observatory, Herzberg Institute of Astrophysics, National Research Council of Canada, 5071 West Saanich Road, Victoria, BC, V9E 2E7, Canada}
\email{sidney.vandenbergh@nrc.gc.ca}

\begin{abstract}

 The Milky Way System and the Andromeda galaxy experienced radically 
different evolutionary histories. Nevertheless, it is found that these two 
galaxies ended up with globular cluster systems in which individual clusters 
have indistinguishable distributions of half-light radii. Furthermore 
globulars in both
M31 and the Galaxy are found to have radii that are independent of their 
luminosities. In this respect globular clusters differ drastically from early-type galaxies in which half-light radius and luminosity are tightly 
correlated. Metal-rich globular clusters in M31 occupy a slightly larger 
volume than do those in the Galaxy.
The specific globular cluster frequency in the Andromeda galaxy is found to he 
significantly higher than it is in the Milky Way System.
The present discussion is based on the 107 Galactic globular clusters, and
200 putative globulars in M31, for which UBV photometry was available.

\end{abstract}


{\it Online material: machine readable tables.}

\section{INTRODUCTION}

    A number of lines of evidence (van den Bergh 2006, Hammer et al.
2007, Brown et al. 2008) indicate that the Milky Way system likely had a 
relatively quiescent formation history, whereas the Andromeda galaxy appears 
to have had a more eventful merger-dominated evolution. It is widely believed 
(e.g. Larsen 2010) that many globular clusters form during the bursts of star 
formation that accompany merger events.
This suggests that the different evolutionary histories of M31 and the Galaxy 
might be reflected in systematic differences between their globular cluster 
systems. Some spice is added to this problem by the fact that the Galaxy may 
have a more massive dark matter halo than M31, even though the Andromeda 
galaxy has a larger stellar mass than does the Milky Way System (Watkins et 
al. 2010). To look into these questions in more detail the present paper 
compares available information on the globular cluster systems in these two 
spiral galaxies.

    Data on the integrated UBV colors of the majority of galactic globular 
clusters have been compiled by Harris (1996) and were updated at 
www.physics.mcmaster.ca/Globular.html. It is the purpose of the present paper 
to provide comparable UBV data for those clusters in the Andromeda galaxy 
which Peacock et al. (2010) have recently flagged as ``old globular clusters''. 
A complete listing of information on these old clusters is given at  
htttp://astro.soton.ac.uk/~m.b.peacock/m31gc.html.  Peacock used the following criteria for includsion of an object in his listing of old globular clusters:  (1) All objects are previously identified in other catalogs, (2) previously published spectroscopy (or HST observations) confirm that they are not background galaxies, (3) colors are consistent with these objects being old clusters.  This age cut is based on g-r color.  Finally (4) preliminary spectral energy distributions obtained so far confirm that most of these objects are genuinely older than 10 Gyr.  Data for these clusters in M31, transformed to the UBV system, are listed in Table 1. Color transformations from the ugr system to the UBV system were 
drawn from Jester et al. (2005) and Peacock et al. (2010).
Values based on color indexes that are uncertain by more than 0.050 mag are 
followed by a colon (:). Applying the transformations used by Jester et al. to 
the colors of globular clusters may introduce slightly larger errors than 
those obtained from objects having the energy distributions of single stars. 
The actual transformation equations used were the
following:

\hspace*{4cm}    V   = g - 0.59(g-r) - 0.01, \hspace*{6cm}               (1)

\hspace*{4cm}    B-V = [(g-r) + 0.22]/ 1.02, \hspace*{6cm}               (2)

\hspace*{4cm}    U-B = [(u-g) - 1.13]/ 1.28. \hspace*{6cm}               (3)
Also given in the table is the reddening-free parameter

\hspace*{4cm}    Q = U-B - 0.72(B-V) \hspace*{7cm}                       (4)
introduced by Johnson \& Morgan (1953). Van den Bergh (1967) has shown that 
this parameter is related to the intrinsic color of a globular cluster by

\hspace*{4cm}     $(B-V)_o$ = Q + 1.00.  \hspace*{7cm}                       (5)

This relation allows one to determine the reddening values, and hence the 
absolute magnitudes, of individual globular clusters in M31. The cluster 
absolute magnitudes and half-light radii listed in Table 1 were calculated 
assuming a distance to M31 of 780 kpc (McConnachie et al. 2005) and using the 
equation

\hspace*{4cm}     $A_v$ = 3.1 E(B-V) \hspace*{7cm}                          (6)\\
   to describe the relation between the total and selective absorption in the 
Andromeda galaxy. For a very small number of objects with negative E(B-V) 
values the foreground absorption was assumed to be zero. All of the objects in 
this class have uncertain (:) photometry.
   While the present paper was being written Wang et al. (2010) published a 
less extensive catalog of UBV data in M31, which was based on the data of 
Galleti et al. (2004, 2006, 2007) and Barmby et al. (2000). A comparison 
between these sets of UBV data shows good agreement. The mean observed differences 
in the sense Wang - van den Bergh are +0.02 $\pm$  0.02 mag in V, -0.07 $\pm$ 0.03  mag in B-V and 0.00 $\pm$ 0.04 mag in U - B.

    Data on 200 putative globular clusters in the Andromeda galaxy, for which 
both half-light radii and UBV data are are available, have been collected in 
Table 1. In Table 2 similar information is provided for 107 Galactic globular 
clusters. The latter data  were drawn from the updated version of Harris 
(1996) and from Mackey \& van den Bergh (2005). The data in both of these 
tables are, of course, most incomplete for intrinsically faint globulars. The 
Andromeda galaxy is seen to contain
16 clusters brighter than $M_{v}$ = -10.0, compared to only 2 such clusters in the 
Milky Way System. The corresponding figures for clusters with
-9.99 $< M_{v} <$ -9.00 and -8.99 $< M_{v} <$ -8.00 are 35 and 10, and 88 and 26, 
respectively. If the luminosity distributions of the globular clusters are 
similar in these two galaxies, then these results indicate that the
M31 cluster system contains about four times as many globular clusters as does 
its Galactic counterpart.

\section{CLUSTER HALF-LIGHT RADII}

    A number of investigations of the dynamical evolution of globular clusters 
(Spitzer \& Thuan 1972, Lightman \& Shapiro 1978, Murphy, Cohn \& Hut 1990, 
Aarseth \& Heggie 1998) have shown that the half-light radii $R_{h}$ of such 
clusters evolve very little over periods as long as ten cluster relaxation 
times. The cluster half-light radius is therefore a useful parameter for the 
study of possible intrinsic differences between the globular clusters in 
different galaxies. In the discussion given below use will be made of the 
median of the half-light radii of various data sets. Beers et al. (1990) have 
shown that this parameter is of comparable accuracy to more complicated 
metrics for populations with  n $>$ 100. The possibility that the radii of 
globular clusters might be a useful parameter for the estimation of distances 
to clusters was first hinted at by Shapley and Sawyer (1927). The fact that 
the radii of globular clusters are in fact (at least within the main body of 
the Milky Way) independent of their luminosities was first established by van 
den Bergh \& Morbey (1984). This conclusion is confirmed by the data on the 
globular clusters in the Andromeda galaxy and in the Milky Way System that are 
discussed below. The result that the half-light radii of globular clusters 
associated with early-type galaxies are independent of globular cluster 
luminosity has also been confirmed over a range of $\sim$2 magnitudes by Jord\'{a}n et 
al. (2005) in the Virgo cluster and, more recently, over a range of $\sim$3 
magnitudes in the Fornax cluster (Masters et al. 2010).

   Figure 1 shows a plot of the $M_{v}$ versus $R_{h}$ relation for the 107 Galactic 
globular cluster in our sample. A similar plot for the 200 putative globular 
clusters in the Andromeda galaxy is shown in Figure 2. Due to observational 
selection effects the M31 cluster sample is highly incomplete for clusters 
fainter than $M_{v}$ $\sim$ -7.
Inspection of Figure 2 hints at the possibility that clusters in the Andromeda 
galaxy with large radii are more common among clusters fainter than $M_v$ = -9.0 than they are 
among more luminous ones. However, a Kolmogorov-Smirnov test shows that this 
apparent effect is only significant at the 95\% level. Figure 1 shows no 
corresponding effect among Galactic globular clusters. The discovery of a few 
very extended globular clusters (Huxor et al. 2008, and references therein) in 
M31 suggests that it would be most useful to discuss the cluster half-light 
radii in terms of their median, rather than in terms of their mean radii. This 
view is supported by the observation (e.g. Harris 2010) that the distribution 
of the effective radii of globular clusters in many galaxies has a tail that 
extends to large radii. Data on the median radii of the globular clusters in 
M31 and in the Galaxy are collected in Table 3. Neither in Andromeda, nor in 
the Milky Way do these data suggest clear-cut evidence for a dependence of 
radii of globular clusters on their luminosity. In this respect globular 
clusters differ radically from early-type galaxies (Giuricin et al. 1988 , 
Nair et al. 2010), in which radii and luminosities are (in all environments)   
tightly correlated. Furthermore, at any given magnitude level, the data in 
Table 3 show no hint for systematic differences between the median radii of 
globular clusters in M31 and in the Galaxy.
Since there appears to be little or no correlation between the radii of 
globular clusters and their half-light radii it is legitimate to inter-compare 
the half-light radii of the total cluster populations in
M31 and in the Galaxy, even though the clusters are sampled to differing 
luminosity limits in these two galaxies. The median half-light radii for these 
two galaxies are found to be: $R_{h}$ = 2.74 pc for 107 globulars in the Galaxy, 
compared to $R_{h}$ = 2.67 pc for 200 putative globular clusters in M31. A Kolmogorov-Smirnov test show no statistically significant difference between the 
distribution of the half-light radii of the clusters in M31 and of those in the Galaxy. It is of particular interest to note that these values appear to be 
indistinguishable from the corresponding values for the globular cluster 
systems surrounding the luminous early-type galaxies in the Virgo and Fornax clusters 
that have recently been studied by Jord\'{a}n et al. (2005) and by Masters et al.
(2010) respectively. The surprising conclusion from these results is that 
globular clusters surrounding galaxies, that were assembled in a wide variety 
of environments, appear to have ended up with similar radii.

    Figure 3 shows a plot of the half-light radii of the M31 globular clusters 
as a function of their projected distance from the nucleus of that galaxy. 
These data show that the radii of clusters increase slightly with increasing 
distance from the nucleus of M31. For 63 globulars with a projected nuclear 
distance $<$ 5.0 kpc the median half-light radius is 2.35 pc, compared to a 
median of 3.11 pc for those 58 clusters that are situated at a projected 
distance $>$ 10.0 kpc from the nucleus of M31. It seems likely that this 32\% 
increase in radius is related to the well-known fact that blue globular 
clusters are, on average, larger than red ones. For example Masters et al. 
(2010) find that the half-light radii of red globular cluster surrounding 
early type galaxies have $<R_{h}>$ = 2.8 $\pm$  0.3 pc , compared to $<R_h>$ = 3.4 $\pm$ 
0.4 pc for blue globular clusters, a 21\% difference.
A caveat is that the results presented above refer to globular clusters of 
above-average luminosity. One therefore cannot yet rule out the possibility 
that different relationships might exist for clusters of below-average 
luminosity.

 \section{LUMINOSITY-RADIUS RELATION}

    Figure 4 shows a plot of the luminosities of globular clusters as a 
function of their projected distance from the nucleus of M31. This figure 
shows that the globular clusters in the central region of the Andromeda galaxy 
appear to be systematically more luminous that those located at greater 
distances from the nucleus of M31. A K-S test shows a 95\% probability that the 
difference between the radial distributions of clusters with $M_{v} <$ -9.0, and 
those with $M_{v} >$ -9.0, is real. Since the present sample contains few clusters 
at $R_{p} <$ 2.5 kpc it appears unlikely that the observed effect is due to 
observational selection against relatively faint clusters in the bright 
central bulge of M31.
Possibly the excess of luminous globular clusters, at small nuclear distances, 
is due to preferential formation of massive clusters in high-density regions. 
Alternatively the most massive clusters might have been dragged inwards by 
tidal friction. A similar plot of $M_{v}$ versus Galactocentric distance $R_{gc}$ for 
the globular clusters in the Milky Way System is shown as Figure 4.8 in the 
monograph by van den Bergh (2000, p.62). Because of differing selection effects 
a direct comparison between the Galaxy and M31 is difficult. In particular it 
is not yet possible to say if M31 shares the excess of intrinsically very 
faint Galactic globular clusters that is seen at $R_{gc} >$ 20 kpc.

 \section{METALLICITY VERSUS Q RELATION}

    The correlation between the published values of the metallicity parameter 
[Fe/H] and the reddening-free index Q for Galactic globular clusters exhibits 
considerable scatter. This is, no doubt, due to (1) observational errors in 
the determinations of [Fe/H], and (2) errors in the observed globular cluster 
colors. Such errors are particularly large for the U-B measurements of faint 
clusters that are superposed on rich star fields. Finally, (3) ``second 
parameter'' effects on the horizontal branches of the color-magnitude diagrams 
of individual globular clusters will introduce an intrinsic dispersion in the 
cluster color versus metallicity relation. Average Q values for different 
metallicity bins are listed in Table 4. The data in the table may be 
approximated by the relation

\hspace*{4cm}    Q $\simeq$~ -0.17 + 0.11 [Fe/H].\hspace*{7cm}               (7)

However, the reader is cautioned that there is no a-priori reason to believe 
that the relation between the metallicity parameter [Fe/H] and the the 
reddening-free color index Q or a globular cluster is, in fact, linear. Spectroscopic data on individual globular clusters in M31, which have a wide range in accuracy, have been published by Perrett et al. (2002).  These data are found to be consistent with Equation 7, which was derived for  Galactic globulars.
Inspection of Figure 5 shows that metal-poor Galactic globular clusters with Q 
$<$ -0.35 occur at all Galactocentric radii.
On the other hand clusters with Q $>$ -0.35 are almost entirely confined to the 
region with $R_{gc} <$ 8 kpc. Finally the four (presumably most metal-rich 
clusters) with Q $>$ -0.05 are all situated at $R_{gc} <$ 3 kpc.
Figure 6 shows a broadly similar situation for the projected nuclear distances 
of clusters in M31. Metal-poor clusters with Q $<$ -0.35 are found at all 
projected nuclear distances, whereas the more metal-rich clusters with Q $>$ -
0.35 all have projected nuclear distances $R_{p} <$ 10 kpc. Finally the three M31 
clusters with the largest Q values, which are presumably the most metal-rich, 
are all situated at $R_{p} <$ 6 kpc. Even though M31 and the Galaxy had very 
different evolutionary histories the radial dependence of metallicity appears 
to be  quite similar in these two galaxies. The projected nuclear distance is 
always smaller than the true nuclear distance.
The data discussed above therefore show that the z\^{o}ne containing metal-rich 
clusters in M31 extends to slightly larger nuclear distances than is the case 
in the Galaxy.

    For the M31 clusters listed in Table 1 the median value of Q = -0.37, 
compared to a median value of Q = -0.35 for for the Galactic globulars in 
Table 7. From Equation 7 one then obtains median values of [Fe/H] = -1.64 and 
[Fe/H] = - 1.82 for the Galaxy and M31, respectvely. 
  A K-S test shows that the distribution of Q values in M31 does not differ, at a statistically significant level, from that of the globular clusters in the 
Galaxy.

 For the M31 clusters listed in Table 1 the median value of Q = -0.37, 
compared to a median value of Q = -0.35 for for the Galactic globulars in 
Table 2. From Equation 2 one then obtains median values of [Fe/H] = -1.64 and 
[Fe/H] = - 1.82 for the Galaxy and M31, respectively.

  A K-S test shows that the distribution of Q values in M31 does not differ at 
a statistically significant level, from that of the globular clusters in the 
Galaxy. However, one should not read too much into this result because both 
the Galactic and M31 samples are strongly biased by the requirement that UBV 
photometry be available. This selection criterion introduces a strong bias 
against metal-rich clusters which are faint (and therefore difficult to 
observe) in the U band of the UBV system.

    It is of some interest to compare the present compilation of reddening 
data in M31 with a similar listing given recently by Fan et al. (2008). Such a 
comparison shows that the reddening values adopted in Table 1 are 
systematically larger by 0.10 $\pm$ 1 0.01 mag than are those obtained by Fan et 
al. The reason for this difference is not yet clear. After correcting for this 
systematic difference the rms difference between the present individual 
reddening values, and those adopted by Fan et al. is 0.08 mag.

\section{X-RAY SOURCES IN M31 CLUSTERS}

    In a recent paper Peacock et al. (2010) have identified 45 X-ray sources 
that appear to be associated with M31 globular clusters.
These authors speculate that high stellar collision rates are the dominant 
factor that determines whether a globular cluster will contain an X-ray 
binary. This hypothesis is strongly supported by the present data which show 
that the 17 M31 globulars in our catalog that contain an X-ray source have a 
median radius $R_h$ = 2.03 pc, which is significantly smaller than the value $R_h$ = 2.67 for all M31 globular clusters in our sample. A K-S test shows that there is only a$\sim$0.01\% chance that the X-ray and non X-ray clusters in M31 were 
drawn from the same parent distribution of sizes. The only object in the 
Peacock et al. sample that does not appear to be compact is the cluster B375 = 
G307, which has a half-light radius $R_h$ = 5.14 pc.
Furthermore, the X-ray clusters in our sample are typically a full magnitude 
brighter than their non X-ray counterparts. A K-S test shows that this 
difference is significant at the 99.5\% confidence level. It is therefore 
concluded that the X-ray clusters in our M31 sample are both smaller, and more 
luminous, than those which do not contain an X-ray source. Both of these 
results support a model in which the collision rate in globular clusters 
determines whether a cluster will contain an X-ray binary.

\section{CONCLUSIONS}

$\bullet$ In both M31 and in the Galaxy the half-light radii of globular
   clusters are found to be independent of their luminosities. In this
   respect globulars differ dramatically from early-type galaxies in
   which radii and luminosities are tightly correlated in all
   environments.

$\bullet$  The median half-light radius of clusters in M31 is found to be
  $R_{h}$ = 2.67 pc in M31, compared to $R_{h}$ = 2.74 pc in the Galaxy.

$\bullet$ In both M31, and the Galaxy, the half-light radii of the metal-poor clusters are found to be systematically larger than those of metal-rich clusters.

$\bullet$ The sizes of metal-poor globular clusters in both the Galaxy and M31 are found to increase with distance from the nucleus.  It follows that the difference between the radii of metal-rich and metal-poor globular clusters is not just due to differing metallicity.

$\bullet$ The distribution of the half-light radii of putative globular
   clusters in M31 is statistically indistinguishable from that of
   the distribution of the radii of Galactic globular clusters.

$\bullet$ In both galaxies the most metal-rich clusters are concentrated
   at small galactocentric distances. However, the region containing
   metal-rich clusters is slightly larger in M31 than it is in the Galaxy.

$\bullet$  At 95\% confidence it is found that the most luminous globulars in
   M31 are more concentrated towards the the center of this galaxy
   than are clusters of lesser luminosity.

$\bullet$ The specific frequency of globular clusters in M31 appears to be three or four times greater than it is in the Galaxy. The excess of M31 clusters
   per unit mass seems to be even greater than this.

$\bullet$ The half-light radii of globular clusters in M31 and in the Galaxy
   seem to be very similar to those of the globulars surrounding
   luminous early-type galaxies in the Virgo and Fornax clusters.
   This suggests that the  radii with which clusters are formed in early-type galaxies are broadly independent of environment.

\section{APPENDIX}

    It is of some interest to compare the radii of Galactic globular clusters 
with those of the globular clusters that are known to be associated with the 
dwarf spheroidal companions of Milky Way system. Listed below in Table 5 are 
data on the half-light radii of the five globular clusters associated with the 
Fornax dwarf, taken from van den Bergh \& Mackey (2004) and information on the 
half-light radii of seven globular clusters which Law \& Majewski (2010) assign with, high or moderate, confidence to the Sagittarius dwarf. The most striking feature of the data in this table is that all 12 of these clusters have radii 
that are larger than the median half-light radius of Galactic globular 
clusters. The median radius of the clusters in Table 5 is $R_h$ = 6.8 pc, which 
is 2.5x larger than the $R_h$ = 2.74 pc median radius of all Galactic globular 
clusters in Table 2. Of the six globulars with $R_{h} >$ 6.8 pc in Table 2, two are 
likely associated with the Sagittarius dwarf. This suggests that: (1) 
satellite clusters of disrupted dwarf companions probably provided a 
significant contribution to total population of large Galactic globular 
clusters, and (2) that only a few mergers with Sagittarius-like dwarfs took 
place during the assembly of the Milky Way system.

    It is a pleasure to thank Mark Peacock for a listing of projected 
galactocentric distances for the globular clusters in M31. I also thank Alan McConnachie for reading the draft manuscript.  I am also indebted 
to Bill and Gretchen Harris for exchanges of e-mails and to Brenda Parrish and 
Jason Shrivell for technical support.

\newpage

\begin{figure}
\plotone{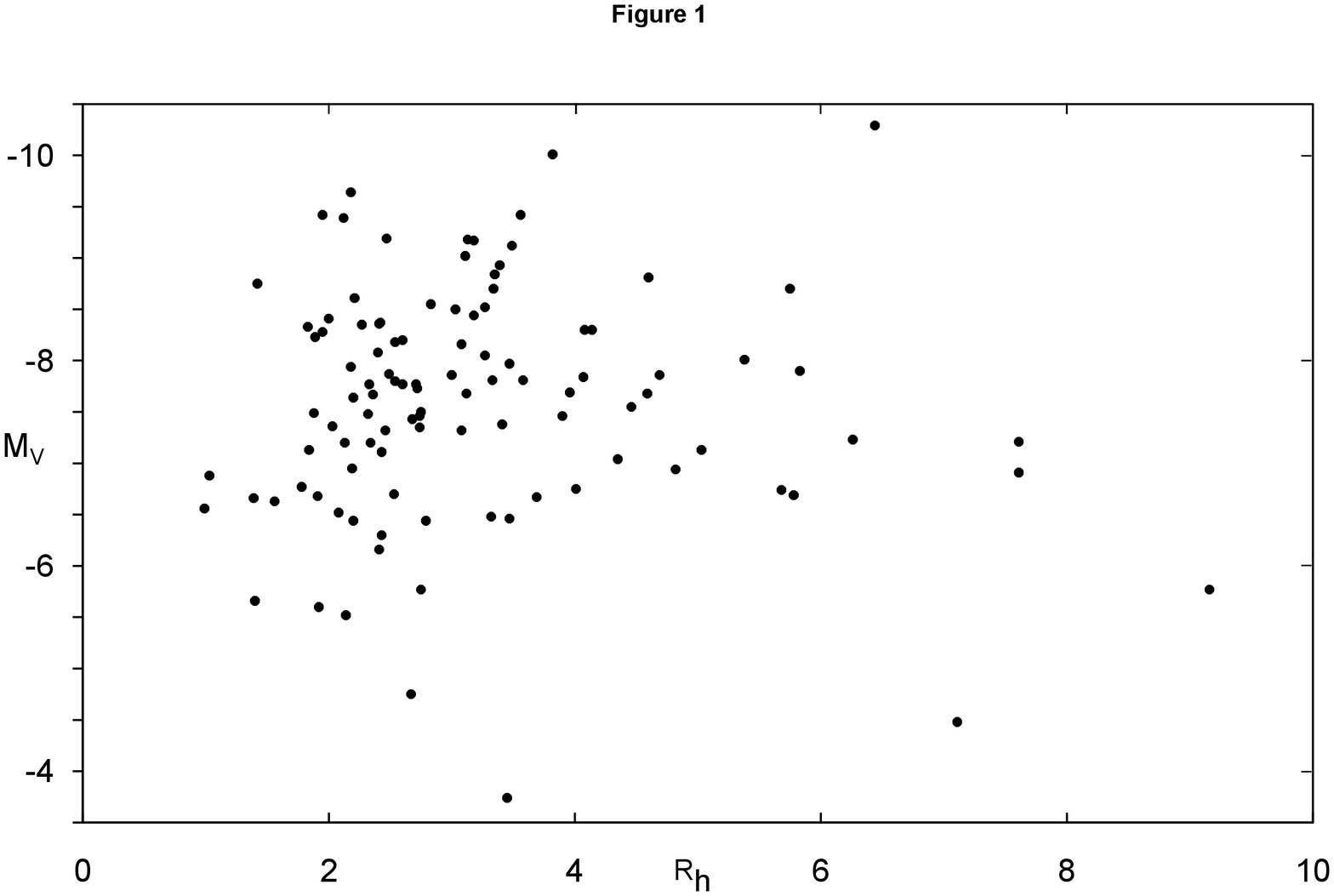}
\caption{Magnitude $M_v$ versus half-light radius $R_{h}$ for
           107 Galactic globular clusters with UBV photometry.
           This plot shows no strong evidence for a correlation
           between globular cluster size and luminosity.}
\end{figure}

\begin{figure}
\plotone{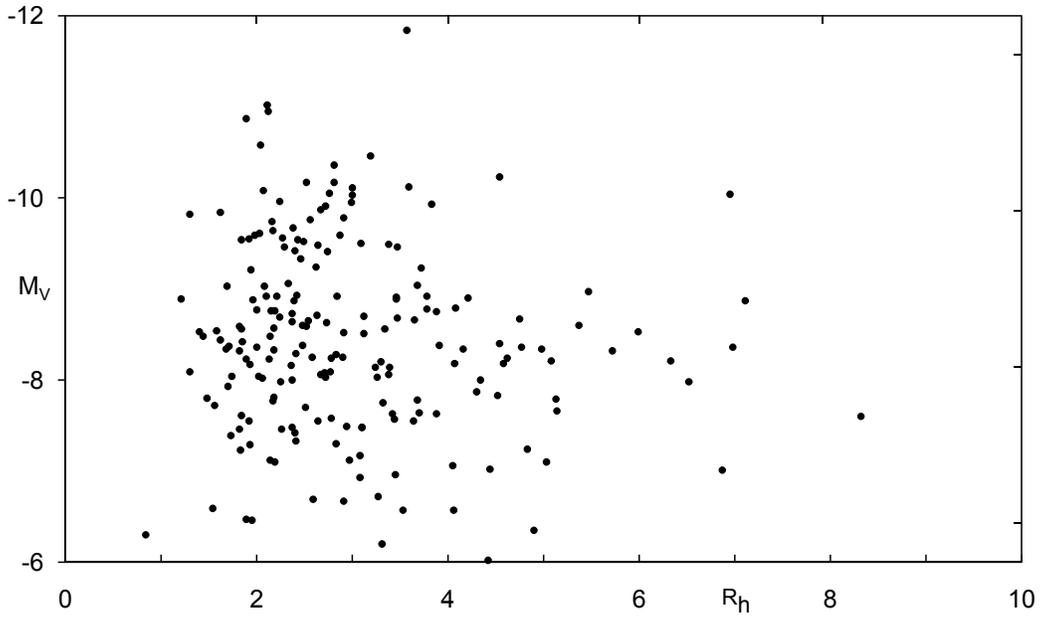}
\caption{Magnitude versus radius half-light radius for putative
           globular clusters in M31 for which UBV photometry
           is available. The data hint at the possibility that
           the clusters at small nuclear distances may be  more
           luminous than those located at greater distances.}
\end{figure}

\begin{figure}
\plotone{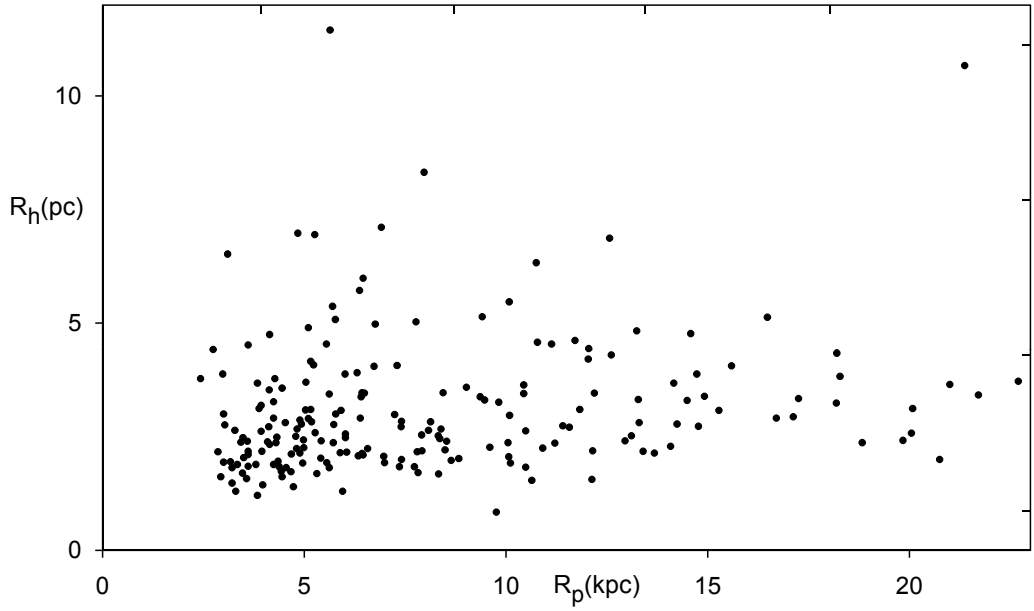}
\caption{Dependence of the half-light radii of globular clusters
           on distance from the center of M31. The outermost
           clusters are, on average, seen to be slightly larger
           than those at smaller radii.}
\end{figure}

\begin{figure}
\plotone{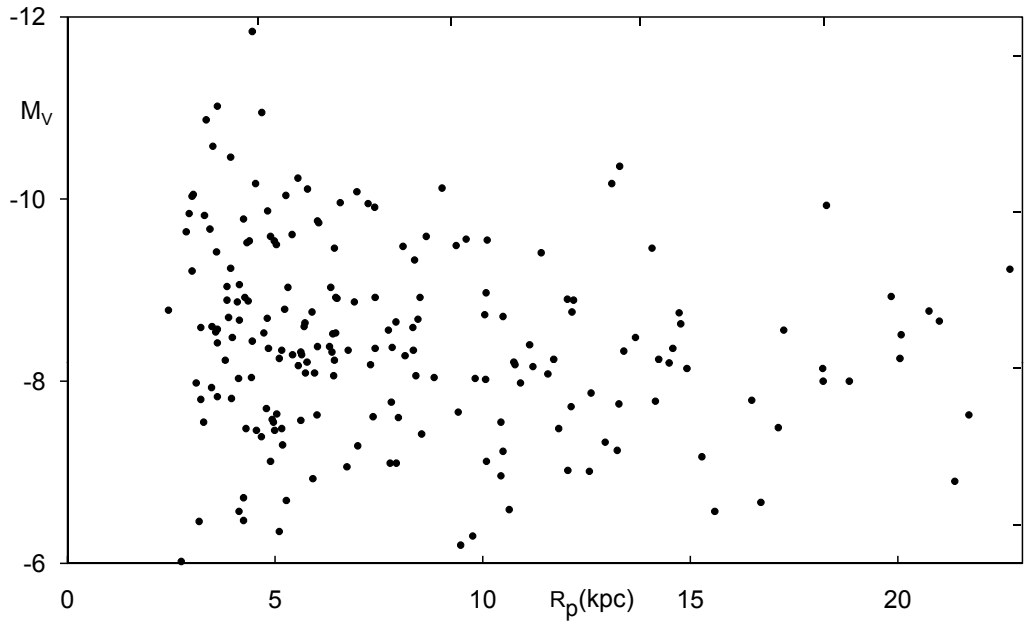}
\caption{Absolute magnitude versus nuclear distance for M31
           clusters. The plot suggests that the most luminous
           clusters are preferentially located at relatively small
           galactocentric distances.}
\end{figure}

\begin{figure}
\plotone{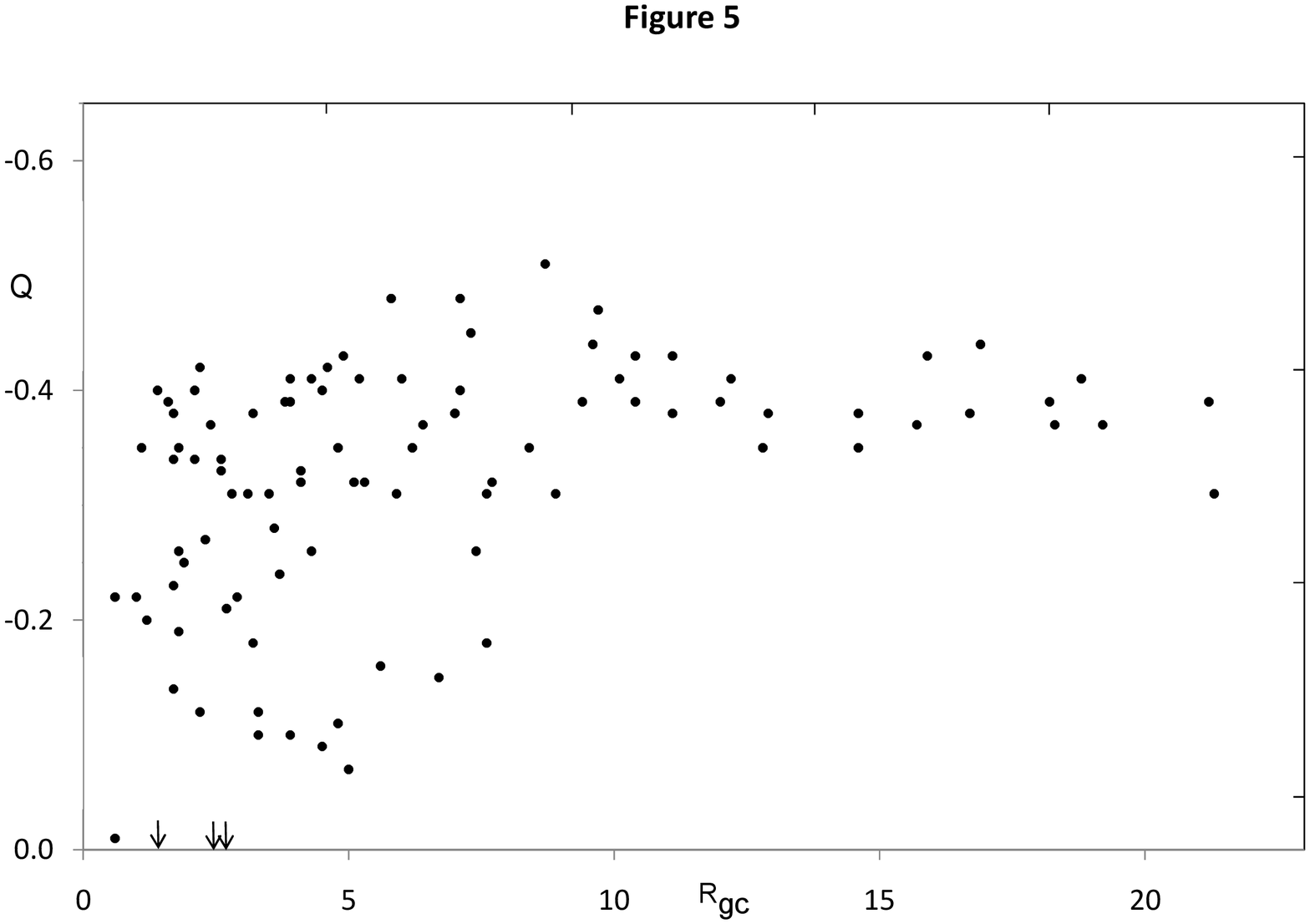}
\caption{Reddening-free parameter Q = U-B - 0.72(B-V)
           versus the Galactic distance for Galactic
           globular clusters. Metal-rich clusters with large
           Q values are seen to occur only at small Galactocentric
           radii.}
\end{figure}

\begin{figure}
\plotone{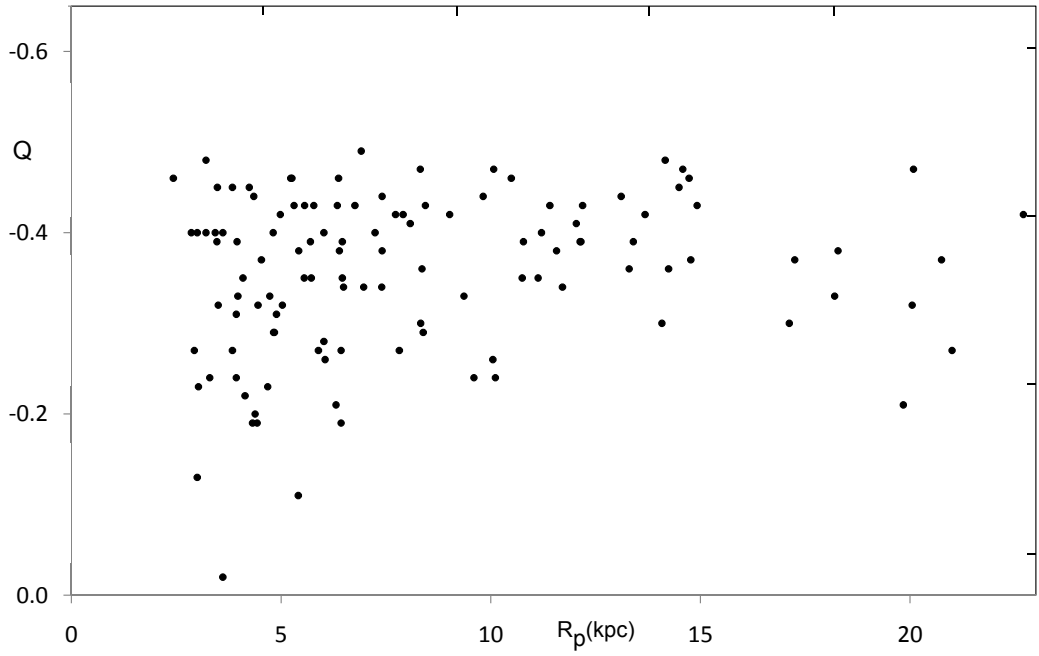}
\caption{Reddening-free parameter Q as a function of projected
           distance Rp from the nucleus of M31. Metal-poor
           clusters occur at all radii, whereas metal-rich
           ones are confined to the central bulge of the
           Andromeda galaxy.}
\end{figure}

\begin{deluxetable}{lcccccccc}
\tablewidth{0pt}
\tablecaption{Data on old clusters in M31}
\tablehead {\colhead{Cluster ID} & \colhead{V} & \colhead{B-V} & \colhead{U-B} & \colhead{Q} & \colhead{E(B-V)} & \colhead{$V_{0}$} & \colhead{$M_{v}$} & \colhead{$R_{h}(pc)$}}

\startdata

B292-G010 & 17.16 & 0.76 & 0.25 & -0.30 & 0.06 & 16.74 & -7.49 & 2.94   \\

B293-G011 & 16.27 & 0.75 & 0.17 & -0.37 & 0.12 & 15.99 & -8.56 & 3.34   \\

B295-G014 & 16.66 & 0.76 & 0.08 & -0.47 & 0.23 & 15.95 & -8.51 & 3.12   \\

B298-G021 & 16.47 & 0.72 & 0.16 & -0.36 & 0.08 & 16.22 & -8.24 & 2.78   \\

B301-G022 & 17.05 & 0.95 & 0.36 & -0.32 & 0.27 & 16.21 & -8.25 & 2.58   \\

B302-G023 & 16.75: & 0.76 & 0.16 & -0.39 & 0.15 & 16.28: & -8.18: & 4.58   \\

B304-G028 & 16.81 & 0.80 & 0.14 & -0.44 & 0.24 & 16.43 & -8.03 & 3.26   \\

B305-D024 & 17.79 & 0.80 & 0.50: & -0.08: & -0.12: & 17.79 & -6.67: & 2.91   \\

B306-G029 & 16.21 & 1.32 & 0.59 & -0.36 & 0.68 & 14.10 & -10.36 & 2.81   \\

B307-G030 & 17.22 & 0.80 & 0.39: & -0.19: & -0.01: & 17.22: & -7.24: & 4.83   \\

B309-G031 & 17.48 & 0.90 & 0.29: & -0.36: & 0.26: & 16.67: & -7.79: & 5.13   \\

B436 & 18.17: & 0.94 & 0.53: & -0.15: & 0.09: & 17.89: & -6.57: & 4.06   \\

B311-G033 & 15.41 & 0.92 & 0.22 & -0.44 & 0.36 & 14.29 & -10.17 & 2.52   \\

B312-G035 & 15.49 & 0.95 & 0.26 & -0.42 & 0.37 & 14.34 & -10.12 & 3.59   \\

B313-G036 & 16.24 & 1.08 & 0.45 & -0.33 & 0.41 & 14.97 & -9.49 & 3.38   \\

B001-G039 & 17.00 & 1.16 & 0.53: & -0.31: & 0.47: & 15.54: & -8.92: & 2.21   \\

B316-G040 & 16.78 & 0.82 & 0.24 & -0.35 & 0.17 & 16.25 & -8.21 & 6.33   \\

B002-G043 & 17.54 & 0.72 & 0.12: & -0.40: & 0.12: & 17.17: & -7.29: & 1.93   \\

B003-G045 & 17.59 & 0.78 & 0.28: & -0.28: & 0.06: & 17.40: & -7.06: & 4.05   \\

B004-G050 & 16.94 & 0.96 & 0.42 & -0.27 & 0.23 & 16.23 & -8.23 & 2.13   \\

B005-G052 & 15.64 & 1.00 & 0.48 & -0.24 & 0.24 & 14.90 & -9.56 & 2.27   \\

B006-G058 & 15.46 & 0.96 & 0.50 & -0.19 & 0.15 & 15.00 & -9.46 & 3.47   \\

B333 & 18.75: & 0.94: & 0.01: & -0.67: & 0.61: & 16.86 & -7.60: & 8.32   \\

B008-G060 & 16.77 & 0.96 & 0.52: & -0.17: & 0.13: & 16.37: & -8.09: & 2.77   \\

B009-G061 & 16.87 & 0.81 & 0.14 & -0.44 & 0.25 & 16.10 & -8.36 & 2.00   \\

B010-G062 & 16.64 & 0.86 & 0.23 & -0.39 & 0.25 & 15.86 & -8.60 & 5.37   \\

B011-G063 & 16.68 & 0.83 & 0.18 & -0.42 & 0.25 & 15.90 & -8.56 & 1.84   \\

B012-G064 & 15.06 & 0.80 & 0.15 & -0.43 & 0.23 & 14.35 & -10.11 & 3.00   \\

H16 & 17.56: & 0.70: & 0.29: & -0.21: & -0.09: & 17.56: & -6.90: & 10.67   \\

B013-G065 & 17.18: & 0.95: & 0.33: & -0.35: & 0.30: & 16.25: & -8.21: & 5.08   \\

B335-V013 & 17.75: & 1.29: & 0.49: & -0.44: & 0.73: & 15.49: & -8.97: & 5.47   \\

B015-V204 & 17.91 & 1.38 & 0.78 & -0.21 & 0.59 & 16.08 & -8.38 & 3.91   \\

B016-G066 & 17.50 & 1.01 & 0.67: & -0.06: & 0.07: & 17.77: & -6.69: & 2.59   \\

DA038 & 18.69 & 0.82 & 0.27: & -0.32: & 0.14: & 18.26: & -6.20: & 3.31   \\

B336-G067 & 17.81 & 0.75 & 0.07: & -0.47: & 0.22: & 17.13: & -7.33: & 2.41   \\

B337-G068 & 16.75 & 0.83 & 0.18 & -0.42 & 0.25 & 15.98 & -8.48 & 2.14   \\

B017-G070 & 15.90 & 1.17 & 0.52 & -0.32 & 0.49 & 14.96 & -9.50 & 3.09   \\

B019-G072 & 14.96 & 0.98 & 0.42 & -0.29 & 0.27 & 14.59 & -9.87 & 2.67   \\

B020-G073 & 14.87 & 0.90 & 0.31 & -0.34 & 0.24 & 14.55 & -9.91 & 2.72   \\

B021-G075 & 17.38: & 1.24: & 0.55: & -0.34: & 0.58: & 16.21: & -8.25: & 2.90   \\

B022-G074 & 17.40 & 0.77 & 0.19: & -0.36: & 0.13: & 17.34: & -7.12: & 2.14   \\

B339-G077 & 16.86 & 0.94 & 0.49: & -0.19: & 0.13: & 16.46: & -8.00: & 2.37   \\

B023-G078 & 14.20 & 1.19 & 0.54 & -0.32 & 0.51 & 12.62 & -11.84 & 3.57   \\

B247 & 17.22: & 0.94: & 0.28: & -0.40: & 0.34: & 16.17: & -8.29 & 11.45   \\

B341-G081 & 16.32 & 0.93 & 0.41 & -0.26 & 0.19 & 15.73 & -8.73 & 2.37   \\

B024-G082 & 16.83 & 0.97 & 0.43 & -0.27 & 0.24 & 16.09 & -8.37 & 1.71   \\

B025-G084 & 16.79 & 1.01 & 0.30 & -0.43 & 0.44 & 15.43 & -9.03 & 1.69   \\

B027-G087 & 15.63 & 0.90 & 0.25 & -0.40 & 0.30 & 14.70 & -9.76 & 2.56   \\

B026-G086 & 17.51 & 1.05 & 0.64: & -0.12: & 0.17: & 16.98: & -7.48: & 2.37   \\

B028-G088 & 16.91 & 0.86 & 0.24 & -0.38 & 0.24 & 16.17 & -8.29 & 2.41   \\

B020D-G089 & 17.53: & 1.16 & 0.44: & -0.40: & 0.56: & 15.79: & -8.67: & 4.75   \\

B029-G090 & 16.68 & 1.07 & 0.66: & -0.11: & 0.18: & 16.12: & -8.34: & 4.16   \\

B032-G093 & 17.59: & 1.22: & 0.56: & -0.32: & 0.54: & 15.92: & -8.54: & 1.58   \\

SK026A & 19.37: & 1.03: & 0.38: & -0.36: & 0.39: & 18.16: & -6.30: & 0.84   \\

B033-G095 & 17.88: & 1.01: & 0.26: & -0.47: & 0.48: & 16.88: & -7.58: & 2.78   \\

B034-G096 & 15.43 & 0.97 & 0.44 & -0.26 & 0.23 & 14.72 & -9.74 & 2.16   \\

B457-G097 & 16.97 & 0.81 & 0.11 & -0.47 & 0.28 & 16.10 & -8.36 & 4.77   \\

B035 & 17.42 & 0.98 & 0.33: & -0.36: & 0.34: & 16.37: & -8.09: & 1.30   \\

B036 & 17.41: & 1.04 & 0.41: & -0.34: & 0.38: & 16.23: & -8.23: & 1.89   \\

B037-V327 & 16.77: & 2.18 & 0.95: & -0.62: & 1.80: & 11.19: & -13.27: & 3.88   \\

B038-G098 & 16.49 & 1.00 & 0.32 & -0.40 & 0.40 & 15.25 & -9.21 & 1.94   \\

B039-G101 & 15.99 & 1.28 & 0.69 & -0.23 & 0.51 & 14.41 & -10.05 & 2.76   \\

B041-G103 & 18.44: & 0.68 & 0.28: & -0.21: & -0.11: & 18.44: & -6.02 & 4.42   \\

B042-G104 & 16.19 & 1.61 & 0.93: & -0.23: & 0.84: & 13.59: & -10.87: & 1.89   \\

B343-G105 & 16.26 & 0.77 & 0.18 & -0.37 & 0.14 & 15.83 & -8.63 & 2.73   \\

B045-G108 & 15.77 & 0.98 & 0.40 & -0.31 & 0.29 & 14.87 & -9.59 & 2.87   \\

B046-G109 & 17.87 & 0.85 & 0.13: & -0.48: & 0.33: & 16.85: & -7.61: & 1.84   \\

B047-G111 & 17.42 & 0.84 & 0.11: & -0.49: & 0.33: & 16.40: & -8.06 & 3.38   \\

B050-G113 & 16.79 & 0.93 & 0.48 & -0.19 & 0.12 & 16.42 & -8.04 & 1.74   \\

SK035A & 19.71: & 1.46: & -0.05: & -1.10: & 1.56: & 14.87: & -9.59: & 1.98   \\

B054-G115 & 18.21: & 0.99 & 0.63: & -0.08: & 0.07: & 17.99: & -6.47: & 1.89   \\

B056-G117 & 17.22 & 1.06 & 0.67: & -0.09: & 0.15: & 16.76: & -7.70: & 2.51   \\

B057-G118 & 17.53 & 0.55 & 0.13: & -0.27: & -0.18: & 17.53: & -6.93: & 3.08   \\

B058-G119 & 14.97 & 0.85 & 0.27 & -0.34 & 0.19 & 14.38 & -10.08 & 2.07   \\

B060-G121 & 16.71 & 0.79 & 0.09 & -0.48 & 0.27 & 15.87 & -8.59 & 1.82   \\

B061-G122 & 16.62 & 1.25 & 0.64: & -0.26: & 0.51: & 15.04: & -9.42: & 2.40   \\

B063-G124 & 15.68 & 1.26 & 0.59 & -0.32 & 0.58 & 13.88 & -10.58 & 2.04   \\

B065-G126 & 16.86 & 0.86 & 0.33 & -0.29 & 0.15 & 16.40 & -8.06 & 2.67   \\

B344-G127 & 15.87 & 0.88 & 0.27 & -0.36 & 0.24 & 15.13 & -9.33 & 2.46   \\

B067-G129 & 17.22 & 0.78 & 0.16 & -0.40 & 0.18 & 16.66 & -7.80 & 1.48   \\

B068-G130 & 16.30 & 1.25 & 0.71 & -0.19 & 0.44 & 14.94 & -9.52 & 2.49   \\

B257-V219 & 17.76 & 1.42 & 0.88: & -0.14: & 0.56: & 16.02: & -8.44: & 1.62   \\

B461-G131 & 17.38 & 0.93 & 0.31: & -0.36: & 0.29: & 16.48: & -7.98: & 2.25   \\

B073-G134 & 15.96 & 0.96 & 0.47 & -0.22 & 0.18 & 15.40 & -9.06 & 2.33   \\

B074-G135 & 16.62 & 0.84 & 0.22 & -0.38 & 0.22 & 15.94 & -8.52 & 2.91   \\

B076-G138 & 16.78 & 0.91 & 0.26 & -0.40 & 0.31 & 15.82 & -9.64 & 2.17   \\

B345-G143 & 16.69 & 0.79 & 0.18 & -0.39 & 0.18 & 16.13 & -8.33 & 2.18   \\

B462 & 18.03 & 0.77 & 0.15: & -0.40: & 0.17: & 17.50: & -6.96: & 3.45   \\

B082-G144 & 15.49 & 1.64 & 1.16 & -0.02 & 0.66 & 13.44 & -11.02 & 2.11   \\

B083-G146 & 17.05 & 0.87 & 0.20 & -0.43 & 0.30 & 16.12 & -8.34 & 4.98   \\

B085-G147 & 16.83 & 0.84 & 0.13 & -0.47 & 0.31 & 15.87 & -8.59 & 2.52   \\

B090 & 18.43 & 0.91 & 0.43: & -0.23: & 0.14: & 18.00: & -6.46: & 1.95   \\

B347-G154 & 16.40 & 0.77 & 0.12 & -0.43 & 0.20 & 15.78 & -8.68 & 3.47   \\

B348-G153 & 16.93 & 0.96 & 0.39 & -0.30 & 0.26 & 16.12 & -8.34 & 1.68   \\

B094-G156 & 15.51 & 0.99 & 0.51 & -0.20 & 0.19 & 14.92 & -9.54 & 1.84   \\

B098 & 16.25 & 0.95 & 0.41 & -0.27 & 0.22 & 15.57 & -8.89 & 1.21   \\

B350-G162 & 16.60 & 0.78 & 0.22 & -0.34 & 0.12 & 16.22 & -8.24 & 4.62   \\

B100-G163 & 17.76 & 0.98 & 0.39: & -0.32: & 0.30: & 16.83: & -7.63: & 3.88   \\

B110-G172 & 15.18 & 0.91 & 0.39 & -0.27 & 0.18 & 14.62 & -9.84 & 1.62   \\

B111-G173 & 16.78 & 0.84 & 0.20 & -0.40 & 0.24 & 16.04 & -8.42 & 1.85   \\

B117-G176 & 16.36 & 0.81 & 0.14 & -0.44 & 0.25 & 15.58 & -8.88 & 1.96   \\

SK052A & 18.41: & 0.84: & 0.10: & -0.50: & 0.34: & 17.36: & -7.10: & 2.19   \\

B351-G179 & 17.55 & 0.85 & 0.15: & -0.46: & 0.31 & 16.59: & -7.87: & 4.30   \\

B352-G180 & 16.49 & 0.78 & 0.10 & -0.46 & 0.24 & 15.75 & -8.71 & 2.63   \\

SK053A & 18.48 & 1.00 & 0.15: & -0.57: & 0.57: & 16.71: & -7.75: & 3.32   \\

B125-G183 & 16.46 & 0.79 & 0.11 & -0.46 & 0.25 & 15.68 & -8.78 & 3.78   \\

DAO55 & 18.74 & 0.77 & 0.04: & -0.51: & 0.28: & 17.87: & -6.59: & 1.54   \\

B354-G186 & 17.74 & 0.79 & 0.23: & -0.34: & 0.13: & 17.34: & -7.12: & 2.97   \\

SK055A & 18.61: & 0.92 & -0.17: & -0.83: & 0.75: & 16.28: & -8.18: & 4.07   \\

B135-G192 & 15.97 & 0.98 & 0.31 & -0.40 & 0.38 & 14.79 & -9.67 & 2.38   \\

B141-G197 & 16.91 & 1.03 & 0.29 & -0.45 & 0.48 & 15.42 & -9.04 & 3.68   \\

B091D-D058 & 15.38 & 1.00 & 0.48 & -0.24 & 0.24 & 14.64 & -9.82 & 1.30   \\

B266 & 18.37: & 1.25 & 0.43: & -0.47: & 0.72: & 16.14: & -8.32: & 1.82   \\

B149-G201 & 16.97 & 1.09 & 0.41: & -0.37: & 0.46: & 15.54: & -8.92: & 3.78   \\

B467-G202 & 17.44 & 0.83 & 0.26: & -0.34: & 0.17: & 16.91: & -7.55: & 3.64   \\

B356-G206 & 16.96 & 0.95 & 0.26 & -0.42 & 0.37 & 15.81 & -8.65 & 2.54   \\

B163-G217 & 14.99 & 1.05 & 0.63 & -0.13 & 0.18 & 14.43 & -10.03 & 3.00   \\

B168 & 17.97: & 1.53 & 0.51: & -0.59: & 1.12: & 14.50: & -9.96: & 2.24   \\

B272-V294 & 18.30: & 1.14 & 0.90: & +0.08: & 0.06: & 18.11: & -6.35: & 4.90   \\

B174-G226 & 15.44 & 1.04 & 0.40 & -0.35 & 0.39 & 14.23 & -10.23 & 4.54   \\

B176-G227 & 16.52 & 0.80 & 0.19 & -0.39 & 0.19 & 15.93 & -8.53 & 5.99   \\

B177-G228 & 18.19 & 0.96 & 0.10: & -0.59: & 0.55: & 16.48: & -7.98: & 6.52   \\

B181-G232 & 16.88 & 0.98 & 0.37: & -0.34: & 0.32: & 15.89: & -8.57: & 2.18   \\

B183-G234 & 15.96 & 1.00 & 0.48 & -0.24 & 0.24 & 15.22 & -9.24 & 2.62   \\

B184-G236 & 17.25 & 1.11 & 0.88: & +0.08: & 0.03: & 17.16: & -7.30: & 2.83   \\

B187-G237 & 17.19 & 1.08 & 0.40: & -0.38: & 0.46: & 15.76: & -8.70: & 3.12   \\

B189-G240 & 17.01 & 1.11 & 0.85: & +0.05: & 0.06: & 16.82: & -7.64: & 3.70   \\

B190-G241 & 16.79 & 0.93 & 0.27 & -0.40 & 0.33 & 15.77 & -8.69 & 2.24   \\

B194-G243 & 17.18 & 0.82 & 0.20 & -0.39 & 0.21 & 16.53 & -7.93 & 1.70   \\

B193-G244 & 15.28 & 1.03 & 0.63 & -0.11 & 0.14 & 14.85 & -9.61 & 2.03   \\

B196-G246 & 17.30 & 0.89 & 0.17: & -0.47: & 0.36: & 16.18: & -8.28: & 2.83   \\

B197-G247 & 17.74 & 1.09 & 0.87: & +0.09: & 0.00: & 17.74: & -6.72: & 3.27   \\

B199-G248 & 17.53 & 0.79 & 0.16: & -0.41: & 0.20: & 16.91: & -7.55: & 1.92   \\

B198-G249 & 17.78: & 0.88 & 0.26: & -0.37: & 0.25: & 17.00: & -7.46: & 1.82   \\

B200 & 18.48 & 1.09 & 0.68: & -0.10: & 0.19: & 17.89: & -6.57: & 3.53   \\

B202-G251 & 17.78: & 0.90 & 0.32: & -0.33: & 0.23: & 17.07: & -7.39: & 1.73   \\

B203-G252 & 16.72 & 0.91 & 0.37 & -0.29 & 0.20 & 16.10 & -8.36 & 6.98   \\

B361-G255 & 16.91 & 0.76 & 0.10 & -0.45 & 0.21 & 16.26 & -8.20 & 3.30   \\

B206-G257 & 15.03 & 0.87 & 0.26 & -0.37 & 0.24 & 14.29 & -10.17 & 2.81   \\

B207-G258 & 17.33 & 0.83 & 0.21 & -0.39 & 0.22 & 16.65 & -7.81 & 2.18   \\

B208-G259 & 17.84 & 1.06 & 0.43: & -0.33: & 0.39: & 16.63: & -7.83: & 4.52   \\

M009 & 17.84: & 0.83 & 0.13: & -0.47: & 0.30: & 16.91: & -7.55: & 2.64   \\

G260 & 16.89 & 0.81 & 0.25 & -0.33 & 0.14 & 16.46 & -8.00 & 4.34   \\

B209-G261 & 16.60 & 0.87 & 0.30 & -0.33 & 0.20 & 15.98 & -8.48 & 1.44   \\

B211-G262 & 16.57 & 0.78 & 0.11 & -0.45 & 0.23 & 15.86 & -8.60 & 2.48   \\

B212-G263 & 15.42 & 0.79 & 0.12 & -0.45 & 0.24 & 14.68 & -9.78 & 2.91   \\

B213-G264 & 16.89 & 0.98 & 0.38 & -0.33 & 0.31 & 15.93 & -8.53 & 1.40   \\

B214-G265 & 17.64 & 0.87 & 0.11: & -0.52: & 0.39: & 16.43: & -8.03: & 2.72   \\

B215-G266 & 17.15: & 0.97 & 0.62: & -0.08: & 0.05: & 17.00: & -7.46: & 2.26   \\

SK083A & 19.43: & 1.16: & 0.03: & -0.81: & 0.97: & 16.42: & -8.04: & 2.02   \\

G268 & 16.51 & 0.96 & 0.42 & -0.27 & 0.23 & 15.80 & -8.66 & 3.65   \\

B217-G269 & 16.46 & 0.93 & 0.32 & -0.35 & 0.28 & 15.59 & -8.87 & 2.39   \\

B218-G272 & 14.65 & 0.90 & 0.34 & -0.31 & 0.21 & 14.00 & -10.46 & 3.19   \\

B219-G271 & 16.41 & 0.96 & 0.42 & -0.27 & 0.23 & 15.70 & -8.76 & 2.15   \\

B363-G274 & 17.82: & 0.73 & 0.07: & -0.46: & 0.19: & 17.23: & -7.23: & 1.83   \\

B220-G275 & 16.57 & 0.83 & 0.14 & -0.46 & 0.29 & 15.67 & -8.79 & 4.08   \\

B221-G276 & 16.78 & 0.96 & 0.34 & -0.35 & 0.31 & 15.82 & -8.64 & 2.37   \\

B224-G279 & 15.23 & 0.80 & 0.12 & -0.46 & 0.26 & 14.42 & -10.04 & 6.95   \\

B279-D068 & 18.45 & 1.08 & 0.51: & -0.27: & 0.35: & 17.36: & -7.10: & 5.03   \\

B225-G280 & 14.13 & 0.97 & 0.47 & -0.23 & 0.20 & 13.51 & -10.95 & 2.12   \\

B228-G281 & 16.78 & 1.02 & 0.38 & -0.38 & 0.40 & 15.54 & -8.92 & 2.84   \\

B229-G282 & 16.40 & 0.77 & 0.06 & -0.49 & 0.26 & 15.59 & -8.87 & 7.11   \\

B230-G283 & 15.99 & 0.75 & 0.11 & -0.43 & 0.18 & 15.43 & -9.03 & 2.08   \\

B365-G284 & 16.61 & 0.83 & 0.14 & -0.46 & 0.29 & 15.71 & -8.75 & 3.88   \\

B231-G285 & 17.28 & 0.89 & 0.21 & -0.43 & 0.32 & 16.29 & -8.17 & 1.93   \\

B232-G286 & 15.63 & 0.81 & 0.16 & -0.42 & 0.23 & 14.92 & -9.54 & 2.43   \\

B233-G287 & 15.82 & 0.86 & 0.21 & -0.41 & 0.27 & 14.98 & -9.48 & 2.64   \\

B281-G288 & 17.63: & 0.93 & 0.39: & -0.28: & 0.21: & 16.98: & -7.48: & 3.10   \\

B234-G290 & 16.82 & 0.96 & 0.41 & -0.28 & 0.24 & 16.08 & -8.38 & 2.48   \\

B366-G291 & 16.15: & 0.78 & 0.15 & -0.41 & 0.19 & 15.56: & -8.90: & 4.21   \\

B255D-D072 & 18.01: & 0.96 & 0.47: & -0.22: & 0.18: & 17.45: & -7.01: & 6.87   \\

B283-G296 & 17.79 & 0.98 & 0.40: & -0.31: & 0.29: & 16.89: & -7.57: & 3.44   \\

B235-G297 & 16.32 & 0.90 & 0.30 & -0.35 & 0.25 & 15.54 & -8.92 & 2.10   \\

B236-G298 & 17.35 & 0.77 & 0.22: & -0.33: & 0.10: & 17.04: & -7.42: & 2.40   \\

B237-G299 & 17.10 & 0.85 & 0.15 & -0.46 & 0.31 & 16.14 & -8.32 & 5.72   \\

B370-G300 & 16.17 & 0.93 & 0.24 & -0.43 & 0.36 & 15.05 & -9.41 & 2.74   \\

B238-G301 & 16.45 & 0.95 & 0.34 & -0.34 & 0.29 & 15.55 & -8.91 & 3.46   \\

B239-M74 & 17.19 & 0.89 & 0.37: & -0.27: & 0.16: & 16.69: & -7.77: & 2.17   \\

B240-G302 & 15.16 & 0.81 & 0.18 & -0.40 & 0.21 & 14.51 & -9.95 & 2.99   \\

B372-G304 & 16.53 & 0.88 & 0.20 & -0.43 & 0.31 & 15.57 & -8.89 & 3.46   \\

B373-G305 & 15.62 & 0.99 & 0.47 & -0.24 & 0.23 & 14.91 & -9.55 & 1.92   \\

SK104A & 17.89 & 0.98 & 0.32: & -0.39 & 0.37: & 16.74: & -7.72: & 1.56   \\

V129-BA4 & 16.98: & 0.82 & 0.19 & -0.40 & 0.22 & 16.30 & -8.16: & 2.36   \\

B375-G307 & 17.57 & 0.98 & 0.44: & -0.27: & 0.25: & 16.80: & -7.66: & 5.14   \\

B378-G311 & 17.60 & 0.77 & 0.12: & -0.43: & 0.20: & 16.98: & -7.48: & 3.10   \\

B382-G317 & 17.34 & 0.82 & 0.12 & -0.47 & 0.29 & 16.44 & -8.02 & 2.06   \\

B386-G322 & 15.56 & 0.88 & 0.33 & -0.30 & 0.18 & 15.00 & -9.46 & 2.29   \\

B327-MVI & 15.91 & 0.80 & 0.16 & -0.42 & 0.22 & 15.23 & -9.23 & 3.72   \\

B391-G328 & 17.19 & 0.88 & 0.25 & -0.38 & 0.26 & 16.38 & -8.08 & 2.71   \\

B393-G330 & 16.87 & 0.91 & 0.31 & -0.35 & 0.26 & 16.06 & -8.40 & 4.54   \\

SK109A & 19.14: & 0.78 & -0.21: & -0.77: & 0.55: & 17.44: & -7.02: & 4.44   \\

B397-G336 & 16.38 & 0.83 & 0.21 & -0.39 & 0.22 & 15.70 & -8.76 & 2.19   \\

B398-G341 & 17.51 & 0.96 & 0.58: & -0.11: & 0.07: & 17.29: & -7.17: & 3.08   \\

B399-G342 & 17.33 & 0.73 & 0.05 & -0.48 & 0.21 & 16.68 & -7.78 & 3.68   \\

B400-G343 & 16.43 & 0.87 & 0.26 & -0.37 & 0.24 & 15.69 & -8.77 & 2.00   \\

B401-G344 & 16.78 & 0.72 & 0.09 & -0.43 & 0.15 & 16.32 & -8.14 & 3.39   \\
B402-G346 & 17.28 & 0.91 & 0.26: & -0.40: & 0.31: & 16.32: & -8.14: & 3.24   \\
BA11 & 17.67 & 0.81 & 0.12: & -0.46: & 0.27: & 16.83: & -7.63: & 3.42   \\
B405-G351 & 15.12 & 0.81 & 0.20 & -0.38 & 0.19 & 14.53 & -9.93 & 3.83   \\
B407-G352 & 16.06 & 0.96 & 0.48 & -0.21 & 0.17 & 15.53 & -8.93 & 2.42   \\
\enddata
\end{deluxetable}

\begin{deluxetable}{lccccc}
\tablewidth{0pt}
\tablecaption{Data on Galactic globular clusters}

\tablehead {\colhead{Cluster ID} & \colhead{[Fe/H]} & \colhead{Q}  & \colhead{$M_{v}$} & \colhead{$R_{h}$} & \colhead{$R_{gc}$}}

\startdata

N104 & -0.76 & -0.26 & -9.42 & 3.56 & 7.4  \\

N288 & -1.24 & -0.39 & -6.74 & 5.68 & 12.0  \\

N362 & -1.16 & -0.39 & -8.41 & 2.00 & 9.4  \\

N1261 & -1.35 & -0.39 & -7.81 & 3.58 & 18.2  \\

Pal 2 & -1.30 & -0.33 & -8.01 & 5.38 & 35.4  \\

N1851 & -1.22 & -0.38 & -8.33 & 1.83 & 16.7  \\

N1904 & -1.57 & -0.41 & -7.86 & 3.00 & 18.8  \\

N2298 & -1.85 & -0.37 & -6.30 & 2.43 & 15.7  \\

N2419 & -2.12 & -0.41 & -9.58 & 17.88 & 91.5  \\

N2808 & -1.15 & -0.38 & -9.39 & 2.12 & 11.1  \\

N3201 & -1.58 & -0.31 & -7.46 & 3.90 & 8.9  \\

N4147 & -1.83 & -0.31 & -6.16 & 2.41 & 21.3  \\

N4372 & -2.09 & -0.48 & -7.77 & 2.60 & 7.1  \\

N4590 & -2.06 & -0.41 & -7.35 & 2.74 & 10.1  \\

N4833 & -1.80 & -0.38 & -8.16 & 3.08 & 7.0  \\

N5024 & -1.99 & -0.37 & -8.70 & 5.75 & 18.3  \\

N5053 & -2.29 & -0.44 & -6.72 & 16.70 & 16.9  \\

N5139 & -1.62 & -0.37 & -10.29 & 6.44 & 6.4  \\

N5272 & -1.57 & -0.41 & -8.93 & 3.39 & 12.2  \\

N5286 & -1.67 & -0.35 & -8.61 & 2.21 & 8.4  \\

N5634 & -1.88 & -0.39 & -7.69 & 3.96 & 21.2  \\

N5694 & -1.86 & -0.42 & -7.81 & 3.33 & 29.1  \\

N5824 & -1.85 & -0.42 & -8.84 & 3.35 & 25.8  \\

N5897 & -1.80 & -0.45 & -7.21 & 7.61 & 7.3  \\

N5904 & -1.27 & -0.35 & -8.81 & 4.60 & 6.2  \\

N5927 & -0.37 & -0.09 & -7.80 & 2.54 & 4.5  \\

N5946 & -1.38 & -0.48 & -7.20 & 2.13 & 5.8  \\

N5986 & -1.58 & -0.35 & -8.44 & 3.18 & 4.8  \\

N6093 & -1.75 & -0.39 & -8.23 & 1.89 & 3.8  \\

N6121 & -1.20 & -0.31 & -7.20 & 2.34 & 5.9  \\

N6101 & -1.82 & -0.43 & -6.91 & 7.61 & 11.1  \\

N6144 & -1.75 & -0.34 & -6.75 & 4.01 & 2.6  \\

N6139 & -1.68 & -0.28 & -8.36 & 2.41 & 3.6  \\

N6171 & -1.04 & -0.10 & -7.13 & 5.03 & 3.3  \\

N6205 & -1.54 & -0.51 & -8.70 & 3.34 & 8.7  \\

N6229 & -1.43 & -0.53 & -8.05 & 3.27 & 29.7  \\

N6218 & -1.48 & -0.40 & -7.32 & 3.08 & 4.5  \\

N6235 & -1.40 & -0.32 & -6.44 & 2.79 & 4.1  \\

N6254 & -1.52 & -0.42 & -7.48 & 2.32 & 4.6  \\

N6256 & -0.70 & -0.19 & -6.52 & 2.08 & 1.8  \\

N6266 & -1.29 & -0.34 & -9.19 & 2.47 & 1.7  \\

N6273 & -1.68 & -0.39 & -9.18 & 3.13 & 1.6  \\

N6284 & -1.32 & -0.31 & -7.97 & 3.47 & 7.6  \\

N6287 & -2.05 & -0.34 & -7.36 & 2.03 & 2.1  \\

N6293 & -1.92 & -0.40 & -7.77 & 2.33 & 1.4  \\

N6304 & -0.59 & -0.12 & -7.32 & 2.46 & 2.2  \\

N6316 & -0.55 & -0.38 & -8.35 & 2.27 & 3.2  \\

N6341 & -2.28 & -0.44 & -8.20 & 2.60 & 9.6  \\

N6325 & -1.17 & -0.35 & -6.95 & 2.19 & 1.1  \\

N6333 & -1.75 & -0.38 & -7.94 & 2.18 & 1.7  \\

N6342 & -0.65 & -0.14 & -6.44 & 2.20 & 1.7  \\

N6356 & -0.50 & -0.18 & -8.52 & 3.27 & 7.6  \\

N6355 & -1.50 & -0.35 & -8.08 & 2.40 & 1.8  \\

N6352 & -0.70 & -0.12 & -6.48 & 3.32 & 3.3  \\

N6366 & -0.82 & -0.07 & -5.77 & 2.75 & 5.0  \\

N6362 & -0.95 & -0.32 & -6.94 & 4.82 & 5.1  \\

N6388 & -0.60 & -0.18 & -9.42 & 1.95 & 3.2  \\

N6402 & -1.39 & -0.33 & -9.12 & 3.49 & 4.1  \\

N6401 & -0.98 & -0.21 & -7.90 & 5.83 & 2.7  \\

N6397 & -1.95 & -0.41 & -6.63 & 1.56 & 6.0  \\

N6426 & -2.26 & -0.38 & -6.69 & 5.78 & 14.6  \\

Ter 5 & 0.00 & +0.14 & -7.87 & 2.49 & 2.4  \\

N6440 & -0.34 & +0.05 & -8.75 & 1.42 & 1.3  \\

N6441 & -0.53 & -0.10 & -9.64 & 2.18 & 3.9  \\

N6453 & -1.53 & -0.26 & -6.88 & 1.03 & 1.8  \\

N6496 & -0.64 & -0.26 & -7.23 & 6.26 & 4.3  \\

N6517 & -1.37 & -0.41 & -8.28 & 1.95 & 4.3  \\

N6522 & -1.44 & -0.22 & -7.67 & 2.36 & 0.6  \\

N6535 & -1.80 & -0.39 & -4.75 & 2.67 & 3.9  \\

N6528 & -0.04 & -0.01 & -6.56 & 0.99 & 0.6  \\

N6539 & -0.66 & -0.31 & -8.30 & 4.08 & 3.1  \\

N6544 & -1.56 & -0.32 & -6.66 & 1.39 & 5.3  \\

N6541 & -1.83 & -0.42 & -8.37 & 2.42 & 2.2  \\

N6553 & -0.21 & +0.09 & -7.77 & 2.71 & 2.2  \\

N6558 & -1.44 & -0.22 & -6.46 & 3.47 & 1.0  \\

I1276 & -0.73 & -0.24 & -6.67 & 3.69 & 3.7  \\

N6569 & -0.86 & -0.22 & -8.30 & 4.14 & 2.9  \\

N6584 & -1.49 & -0.38 & -7.68 & 3.12 & 7.0  \\

N6624 & -0.44 & -0.20 & -7.49 & 1.88 & 1.2  \\

N6626 & -1.45 & -0.32 & -8.18 & 2.54 & 7.7  \\

N6638 & -0.99 & -0.27 & -7.13 & 1.84 & 2.3  \\

N6637 & -0.70 & -0.25 & -7.64 & 2.20 & 1.9  \\

N6642 & -1.35 & -0.23 & -6.77 & 1.78 & 1.7  \\

N6652 & -0.96 & -0.31 & -6.68 & 1.91 & 2.8  \\

N6656 & -1.64 & -0.43 & -8.50 & 3.03 & 4.9  \\

Pal 8 & -0.48 & -0.16 & -5.52 & 2.14 & 5.6  \\

N6681 & -1.51 & -0.40 & -7.11 & 2.43 & 2.1  \\

N6712 & -1.01 & -0.31 & -7.50 & 2.75 & 3.5  \\

N6715 & -1.58 & -0.37 & -10.01 & 3.82 & 19.2  \\

N6717 & -1.29 & -0.37 & -5.66 & 1.40 & 2.4  \\

N6723 & -1.12 & -0.33 & -7.84 & 4.07 & 2.6  \\

N6749 & -1.60 & -0.74 & -6.70 & 2.53 & 5.0  \\

N6752 & -1.56 & -0.41 & -7.73 & 2.72 & 5.2  \\

N6760 & -0.52 & -0.11 & -7.86 & 4.69 & 4.8  \\

N6779 & -1.94 & -0.47 & -7.38 & 3.41 & 9.7  \\

N6809 & -1.81 & -0.41 & -7.55 & 4.46 & 3.9  \\

N6838 & -0.73 & -0.15 & -5.60 & 1.92 & 6.7  \\

N6864 & -1.16 & -0.35 & -8.55 & 2.83 & 14.6  \\

N6934 & -1.54 & -0.35 & -7.46 & 2.74 & 12.8  \\

N6981 & -1.40 & -0.38 & -7.04 & 4.35 & 12.9  \\

N7006 & -1.63 & -0.48 & -7.68 & 4.59 & 38.8  \\

N7078 & -2.26 & -0.43 & -9.17 & 3.18 & 10.4  \\

N7089 & -1.62 & -0.39 & -9.02 & 3.11 & 10.4  \\

N7099 & -2.12 & -0.40 & -7.43 & 2.68 & 7.1  \\

Pal 12 & -0.94 & -0.43 & -4.48 & 7.11 & 15.9  \\

Pal 13 & -1.74 & -0.66 & -3.74 & 3.45 & 26.7  \\

N7492 & -1.51 & -0.07 & -5.77 & 9.16 & 24.9 \\

\enddata
\end{deluxetable}

\begin{deluxetable}{cclcc}
\tablewidth{0pt}
\tablecaption{Median radii of M31 and Galaxy clusters}

\tablehead {&\colhead{Andromeda} & &  \colhead{Milky Way}\\

 \colhead{$M_v$}  & \colhead{$R_h$} & \colhead{n}  & \colhead{$R_h$} & \colhead{n}\\
 & \colhead{(pc)}  &    & \colhead{(pc)}}

\startdata

 $<$ -10.00    &  2.9 &  16 &    ... &  2\\
-9.00 to -9.99 &  2.4 &  35 &    3.1  &  10\\
-8.00 to -8.99 &  2.7 &  88 &    2.9  &  26\\
-7.00 to -7.99 &  2.9 &  45 &    2.7  &  39\\
-6.00 to -6.99 &  3.2 &  15 &    2.4  &  21\\
   $>$ -6.00    &   ... &  0  &   2.4:  &   7\\

\enddata
\end{deluxetable}

\begin{deluxetable}{cc}
\tablewidth{0pt}
\tablecaption{Color-metallicity relation for Galactic globular clusters}
\tablehead {\colhead{[Fe/H]} & \colhead{$<Q>$}}

\startdata

0.00 to  -0.49 &   0.23  $\pm$  0.07\\
-0.50 to -0.99 &   0.22  $\pm$   0.02\\
-1.00 to -1.49 &   0.34  $\pm$   0.02\\
-1.50 to -1.99 &   0.39  $\pm$   0.01\\
-2.00 to -2.49 &   0.39  $\pm$  0.02\\

\enddata
\end{deluxetable}

\begin{deluxetable}{ll}
\tablewidth{0pt}
\tablecaption{Half-light radii of globular clusters associated with the Fornax and Sagittarius dwarf systems}
\tablehead {\colhead{Name} & \colhead{$R_h$}}

\startdata

Fornax 1 & 11.8 pc\\
Fornax 2 & 8.2\\
Fornax 3 & 4.4\\
Fornax 4 & 3.5\\
Fornax 5 & 4.4\\
NGC 5053 & 16.7\\
NGC 5634 & 4.0\\
NGC 6715 & 3.8\\
Arp 2 & 15.9\\
Palomar 12 & 7.1\\
Terzan 7 & 6.6\\
Terzan 8 & 7.6\\

\enddata
\end{deluxetable}

\end{document}